\documentclass[authoryear,preprint,12pt]{article}
\usepackage{t1enc}
\usepackage[latin1]{inputenc}

\usepackage{amsmath}
\usepackage{amssymb}

\usepackage{colortbl}

\usepackage[T1]{fontenc}
\usepackage{times}
\usepackage{graphicx}

\newcommand{\ds}{\displaystyle}

\title{Impact of Risk of Storm on Faustmann Rotation} 
\author{ Patrice
 Loisel\thanks{ INRA, UMR 729 MISTEA, F-34060 Montpellier, France}
}
\date{}

\begin{document}
\hyphenation{le-vels} 
\hyphenation{res-pec-ti-ve-ly}
\hyphenation{re-fe-ren-ce} 
\hyphenation{A-ve-ra-ge} 
\hyphenation{Mo-dels}
\hyphenation{a-bi-li-ty} 
\hyphenation{pa-ra-me-ters}
\hyphenation{ac-tua-li-zed} 
\hyphenation{cha-rac-te-ris-tics}
\hyphenation{re-plan-ting}
\hyphenation{des-cri-bed}
\hyphenation{asso-cia-ted}
\hyphenation{stu-dies}
\hyphenation{com-pa-ra-tive}
\hyphenation{vo-lu-me}
\hyphenation{va-ria-ble}
\hyphenation{de-pre-cia-tion}
\hyphenation{dif-fe-ren-tia-ted}
\hyphenation{in-te-rest}
\hyphenation{stu-died}
\hyphenation{va-rious}

\maketitle

{\bf Highlights}
\begin{small}
\begin{itemize}
\item  Modeling the impact  of storm on forest damage and depreciation of timber price.
\item The greater the discount rate, the earlier the optimal thinnings are favored.
\item With risk, the optimal thinnings are made earlier than in the case without risk.
\item We analyze joint influence of storm risk and criteria on optimal silviculture.
\end{itemize}
\end{small}

{\bf Abstract} 

Global warming may induce in Western Europe an increase in
storms. Hence the forest managers will have to take into account the
risk increase. We study the impact of storm risk at the stand
level. From the analytical expressions of the Faustmann criterion and
the Expected Long-Run Average Yield, we deduce in presence of storm
risk the influence of criteria and of discount rate in terms of
optimal thinnings and cutting age. We discuss the validity of using a
risk adjusted discount rate (a rate of storm risk added to the
discount rate) without risk to mimic the storm risk case in terms of
optimal thinnings.


\section{Introduction}

Global warming may induce in Western Europe an increase in storms
(\cite{Haarsma}) and also a modification of the distributions
governing their frequency and severity. Wind storms will induce high
yield losses { for forest managers in terms of timber losses and
  clearing costs (\cite{Hanewinkel2013}). Moreover, after a storm, due
  to an influx of wood on the market, the timber price and hence the
  future value of forest will decrease}.  So the forest managers will
have to take into account the risk increase { and its consequences for
  stand forest}.  Therefore they will probably have to modify the
rotation period and more generally the silviculture.

{ In the absence of risk, \cite{Faust} proposed a formalism, based on
  the expected discounted income, which allows to determine the
  optimal rotation period}.  Many authors have been studying
(\cite{Naslund}, \cite{Schreuder}, \cite{Clark}, \cite{KaoBrodie}, $\
$ \cite{HaightMonserudChew}) the determination of optimal thinning and
cutting age at the stand level. In parallel, empirical studies
examined the economic impact on optimal silviculture: $\ \ \ \ \ \ \ \ \ \ \ \ $
\cite{Brodie} analyzed the impact of discount rate on optimal
thinnings by simulation for Douglas stand, \cite{Hyytiainen} studied
the joint influence of the rate of interest and the initial state on
the rotation period for Spruce and Scots Pine sites.

The risk of destruction has been introduced for forest stands by 
\cite{Mart} and $\ $ \cite{Rout} in discrete time. Thereafter, \cite{Reed}
has studied the optimal forest rotation in continuous time with the
risk of fire.  \cite{Thor} analyzed endogenous risk,
More recently { concerning natural risk}, \cite{Stau} studied the
impact of risk on the expected value of a Spruce stand for various
hazard rate functions,
\cite{Loisel} examined the impact of density dependence growth on
optimal cutting age.{ \cite{Price}
    focused on the validity of using the rate of physical risk, added
    to the discount rate as a new adjusted discount rate}. 

  More precisely, concerning the risk of storm, \cite{Hanew} studied
  the impact of storm on the stand forest.
  \cite{Holecy} analyzed model insurance. But few works in the
  literature were focused on the impact of storm risk on forest
  rotation: $\ $ \cite{HaightSmithStraka} studied the impact of storm on
  the expected present value without taking thinnings into
  account. \cite{Meilby} focalized their analysis on shelter effect to
  prevent windthrow in multiple-stand model { but they considered
    an exogeneous land value}. Moreover, using empirical material
  \cite{Deegen} studied the combined effects if storm survival
  probabilities and site productivities change
  simultaneously. \cite{Chang} evaluated the impact of
  hurricane-related production risk in Pine plantations using a
  generalized Reed model. { In all these numerical studies
    concerning the storm risk,
    thinnings are either fixed or not taken into account. There is a
    lack of studies permitting the analysis and the prediction of the
    modification induced by storm risk on optimal thinnings and
    cutting age}.
  { In contrast with the previous studies, our analysis is generic
    and based on the analytic expressions of the criteria. More
    precisely, it uses the relative contribution of thinnings income
    in case of storm risk.} In this paper, the analytical nature of
  the proposed methology is a novelty in contrast with the previous
  works available in the literature, which were based on
  empirical/numerical techniques.

  In the present work, we model the impact of storm risk on optimal
  silviculture at the stand level.
We consider two criteria { which permit to evaluate scenarii with
  various thinnings and cutting age}: the Expected Discounted Value
(the Faustmann criterion) and the Expected Long-Run Average Yield. We
optimize these criteria in presence of storm risk. We consider an
adaptation of the model of \cite{Reed}
  towards a forest stand to take into account thinnings for the
  specific storm risk case.
Moreover, the depreciation of timber price due to an influx of wood on
the market after a storm is considered. The goal is to determine the
joined influence of the presence of storm risk and of the chosen
maximized criteria on optimal thinnings and rotation period.

In the first section, we consider { the reference case without the
  storm risk, which will be used for comparison}, In the second section,
we present the different models in case of storm risk: the storm risk
process, the impact of the storm on the stand forest, the timber price
depreciation.  Assuming we know the expected thinning incomes, final
income at cutting age and the various costs link to storm risk, we
develop the obtaining of analytical expressions of the Faustmann Value
and the Expected Long-Run Averaged Yield in the presence of storm risk
{ taking the depreciation of timber price reference into account}.
From these analytic expressions, we highlight in presence of storm
risk the influence of criteria and of discount rate in terms of
optimal thinnings and cutting age. Even if, the influence of the
discount rate or the level of risk has been previously observed
empirically in specific forest stands, the originality of our work is
to provide an explanation of generic properties: the obtained results
do not depend on the species or the forest growth. Moreover we discuss
the validity of using a risk adjusted discount rate (a rate of storm
risk added to the discount rate) without risk to mimic the storm risk
case in terms of optimal thinnings. Due to the fact that, if a storm
occurs at a date anterior to a fixed time limit, the storm has no
impact, we infer that the time limit is an important threshold.  The
relative positions in time of the optimal thinning without risk and of
the time limit allow to deduce the behavior of the optimal thinning
with respect to the risk. Finally we illustrate and confirm the
obtained economic conditions by considering a beech stand.

\section{In absence of storm risk}

We first consider a forest stand in the absence of storm risk. The
analysis of this case will allow us { to give the notations and} to
define a benchmark management of the stand, useful for comparison with
the more complex case in presence of storm risk.

\subsection{The Faustmann value}

For a cutting age $T$, a discount rate $\delta$, the Faustmann Value
$J_0$ taking into account thinning incomes (up to the constant cost of
regeneration $c_1$) of a stand is the discounted value of cutting
incomes minus cost of regeneration $c_1$:
\begin{align}
 J_0 =  \sum_{i=1}^{+\infty} (W(0,T)-c_1) e^{- i \delta T} =
{(W(0,T)-c_1) e^{-\delta T} \over 1- e^{-\delta T}} =
{W(0,T)-c_1\over e^{\delta T} -1}
\end{align}
where $W(0,T)$ is the total income on $[0,T]$ composed of the sum of
thinning incomes summed on $[0,T]$ actualized at time $T$ and the
final income.

We express the thinning income and the final income. Let $N$ the
number of thinning dates, let the thinning dates $(u_k)_{k=1..N}$ such
that $0 < u_1 < u_2 < .. < u_N < T$ and $h_k$ the vector of the
corresponding rate of thinnings at these dates. Let a timber price
reference $p_0$, which can depend on the conjoncture, {$R(p_0,t)$ the
  vector of potential income at time $t$.} We denote $R_k=
R(p_0,u_k)$. 
Hence $\mathcal H_0(t_1,t_2)$ the thinning income on period
$[t_1,t_2]$ actualized to time $t_2$ is given by: $\ds \mathcal
H_0(t_1,t_2) = \sum_{t_1 < u_k \leq t_2}^N {\bf \ ^tR_k.h_k} \
e^{\delta (t_2-u_k)} $ and the final income: $\ds V(T)={\bf \
  ^tR(p_0,T).1}$. Thus the total income is:
\begin{align}
 W(0,T)= \mathcal H_0(0,T) + V(T)= \sum_{k=1}^N {\bf \
  ^tR_k.h_k} \ e^{\delta (T-u_k)} + V(T)
\end{align}
 { The expression of income
  $R(p_0,t)$ depends on the type of the model chosen for forest
  growth: $R(p_0,t)$. For a stand model \cite{Clark} or an average
  tree model, $h_k$ is reduced to scalar. For a size-structured model
  or a tree-individualized model, the components of $h_k$ are related
  to the tree-sizes, classically the tree-basal area}. { We assume
  that income $R$ is linear in its first argument}.

\subsection{The Long Run Average Yield}

The Long-Run Average Yield is defined by the ratio of the averaged net
economic return (stumpage minus regeneration cost) and the cutting age:
\begin{align}
 \overline Y_0 = { W_0(0,T) -c_1 \over T} \ \mbox{ where } \ W_0(0,T) =
\sum_{k=1}^N {\bf \ ^tR_k .h_k} + V(T)
\end{align}

\subsection{Maximization of the Faustmann Value}

We consider the maximization of the Faustmann Value with respect to
thinnings and cutting age: $\ds \max_{(h_k)_k,(u_k)_k,T} J_0$.  For
fixed cutting age $T$, the maximization of the Faustmann Value $\ds
J_0$ is equivalent to the maximization of $W(0,T)$. Hence the
maximization with respect to thinnings and cutting age can be
decomposed into two levels: $\ds \max_T \ds [W^*(0,T) -c_1] /
(e^{\delta T} -1)$ where $\ds W^*(0,T)$ is the maximum of $W(0,T)$
with respect to thinnings: $\ds \max_{(h_k)_k,(u_k)_k} W(0,T)$.

For a fixed cutting age $T$, the behavior of the coefficient relative to
the thinning income $\ds \beta^k_{\delta,0} = e^{\delta(T-u_k)}$ in
the income $W(0,T)$ be studied in order to deduce the influence of the
discount rate in terms of optimal thinnings.

\subsubsection*{Dependence of optimal thinning with respect to the
  discount rate}

The derivative of $\beta^k_{\delta,0}$ with respect to the discount
rate $\delta$ normalized by $\beta_{0,0}^k$ is:
\begin{align}
 {1 \over \beta_{0,0}^k} {\partial \beta_{\delta,0}^k \over \partial
  \delta}|_{\delta= 0} = T-u_k 
\end{align}
We deduce that the relative additional contribution of $\ds \beta_{\delta,0}^k$
decreases as $k$ increases.
Hence for a fixed cutting age $T$, the greater the discount
rate $\delta$, the earlier the optimal thinnings.
{ Moreover, the optimal cutting age decreases with respect to the
  discount rate (Appendix A). Hence, also for the optimal cutting age,
  by considering a fixed final tree-density and a fixed number of
  thinning dates (only thinning dates can be changed), the greater the
  discount rate $\delta$, the earlier the optimal thinnings}.

This result is not surprising and well known: the greater the discount
rate, the lower the actualized income for the last thinnings.
In earlier works such as in \cite{Brodie}, a similar result has been
checked by simulation for specific Douglas stand. In contrast to 
this study, 
the obtained result is generic and does not depend on the forest growth.

 \section{In the presence of storm risk}

 In order to deduce the behavior of a forest in presence of storm
 risk, we present the risk model, then obtain analytical expressions
 of the criteria that permit a comparison of the optimal silviculture
 without and with risk.

 \subsection{The storm risk modeling}

The storm risk modelling is based on various models which interact: the
 model of the risk of storm, the model of the impact of a storm on a
 stand forest and finally the model of timber price depreciation
 following a storm.

 \subsubsection*{Model of the storm risk}
 
 As in \cite{Reed}, we assume that storms occur in a Poisson process
 i.e. that storms occur independently of one another, and randomly in
 time.  Thus the distribution of the times between successive storms
 $\tau$ is an exponential with mean $\ds 1 / \lambda$: $\ds F(x) = 1-
 e^{-\lambda x}$ where $\lambda$ is the expected number of storms per
 unit time. The severity of the storm is given by the random variable
 $\mathcal A$.  The storm risk is described by the couple of random
 variables $(\tau,\mathcal A)$.

 \subsubsection*{Model of the impact of a storm on the stand forest}

 The consequences of storm are characterized by the fact that the
 impacts are very differentiated according to the stand age at which
 the storm occurs. The storms have a low impact on young stands
 (\cite{Hanew}):
 for a tree-height $H$ less or equal than $H_L$, there is no damage
 for the trees. The height $H_L$ is reached at a time $t_L$, the time
 $t_L$ explicitly depends on the height limit $H_L$ and can be deduced
 using the time dependent function $H$: $H(t_L)=H_L$. Therefore, in
 case of storm at time $\tau \leq t_L$ the storm has no impact,
for a storm occuring at time $ \tau > t_L$ the proportion of damaged
trees $\theta_t$ is positively correlated with the severity $\mathcal
A$ of the storm.  Moreover, following \cite{Hanew}, the proportion of
damaged trees $\theta_t$ depends on the state of the stand at the date
$\tau$ the storm occurs: height $H(\tau)$, tree-diameter $d(\tau)$.
Let $\alpha(t)$ the expectation of the proportion of surviving trees
$1-\theta_t$. 
It is possible to integrate specific behavior for $\theta$ hence for
$\alpha(\tau)$, such as the dependence on the height $H(\tau)$ and the
$\ds H(\tau) / d(\tau)$ ratio or the time spent since last thinning,
relative thinning volume in the latest thinning (\cite{Lohmander}).
{ Tree damages increase with respect to the $H/d$ ratio. In order
  to prevent tree-damages, forest managers make earlier thinnings.}

\subsubsection*{Model of the timber price depreciation}

Due to a timber influx on the market following the storm, we observe a
variable timber price depreciation. $\xi_t$ the rate depreciation
of timber price is positively correlated with the severity $\mathcal
A$. Hence, generally $\theta$ and $\xi$ are positively correlated
random variables. We denote $\alpha_d$ the expectation of $\xi$.
We also denote $\alpha_p(t)$ the expectation of $(1-\theta_t)(1-
\xi_t)$.
To simplify, we assume here $\xi$ constant, so $\xi \equiv \alpha_d$,
hence: $\alpha_p(t) = (1-\alpha_d) \alpha(t)$.
We take into account a lasting impact of the last storm, occuring
before time $t_L$, on the timber price. Let $\tau^-_t < t_L$ the date
of the last storm before $t$ and $\rho_{\tau^-_t}(t)$ the rate
depreciation of timber price at time $t$.  The timber price reference
$p_0$ in the case without risk is then replaced by
$p_t=(1-\rho_{\tau^-_t}(t))p_0$ in the presence of storm risk.
The rate depreciation $\rho_{\tau^-_t}(t)$ decreases with respect to
time $t$ and only depends on the elapsed time since the last storm:
$t-\tau^-_t$, hence it is sufficient to define $\rho_0(t)$ and then
deduce $\rho_{\tau^-_t}(t) = \rho_0(t-\tau^-_t)$. $\rho_0(.)$
decreases with respect to the elapsed time since the last storm with
$\rho_0(0)=\alpha_d$.
The support of $\rho_0(.)$ is $[0,\Delta]$ 
where $\Delta$ is the perturbation time of impact for the rate
depreciation of the timber price.
Finally, if the last storm before $t$ occurs at time $\tau_t^-$, the
timber price reference $p_t$ is given by:
\begin{align}
  \label{price}
  p_t = (1-\rho_0(t-\tau^-_t))p_0 & 
\end{align}
For example, with a linear increasing of the timber price: $\ds
\rho_0(t) =\alpha_d (1-{t \over \Delta})$ for $ 0 \leq t < \Delta $.
We notice that the date of the last storm $\tau^-_t$ is a random
variable, hence the rate depreciation and the reference price are
random variables.
 
\subsection{The Faustmann Value}

  Let $\tau$ the spending time between the beginning of the stand and
  the first event of the stand after $t_L$, either by storm or by
  logging at time $T$. The distribution of the spending time $\tau$ is:
\begin{align}
 F_{\tau}(t) = F(t-t_L) = 1-e^{-\lambda (t-t_L)} \mbox{ for } t_L < t \leq T.
\end{align}
We consider an adaptation of the simplified scenario of
   \cite{Reed} to our case of storm risk:

   - if a storm occurs at time $\tau \leq t_L$, the storm has no
   impact, { the stand continues to grow.}

   - if a storm occurs at time $\tau$ with $t_L < \tau \leq T$, the
   proportion of damaged trees $\theta_t$ is { non neglectable}, a
   clearcutting and a regeneration of the stand is made at time
   $\tau$.

   - if no storm occurs before time $T$, a clearcutting and a
   regeneration of the stand is made at time $T$.

   We assume that perturbation time $\Delta$ is sufficiently small
   such that $\Delta < \min(t_L,T-t_L)$ and the timber depreciation
   have no impact on the thinning incomes before time $\Delta$: for
   small tree-diameter the conjoncture does not modify timber price.
   These conditions are commonly checked and avoid a link (in terms of
   the price reference) between successive rotations.
 
   Let $\tau_i$ the spending time between the beginning of the stand
   and the first event of the stand after $t_L$, either by storm or by
   logging at time $T$ for the $i$th rotation. The net economic return
   $\mathcal Y_i$ (thinning incomes, final income minus costs)
   actualized at final time $\tau$ is given by:
\begin{equation}
  \mathcal Y_i =
\begin{cases}
  \mathcal H(0,\tau_i,\sigma) +
  \mathcal V(\theta_\tau,\xi_\tau,\tau_i) -c_1-
  C_n(\theta_\tau,\tau_i) & \mbox{ if }  t_L < \tau_i < T \\
  \mathcal H(0,T,\sigma) +V(T)-c_1 & \mbox{ if } \tau_i = T
\end{cases}
\end{equation}
where $\ds \mathcal H(0,\tau_i,\sigma)$ is the thinning income (taking
into account the previous storm $\sigma$, see Appendix B.1), $\ds
\mathcal V(\theta_\tau,\xi_\tau,\tau)$ is the final income,
$C_n(\theta_\tau,\tau)$ are the clearing costs
for a storm occuring at time $\tau < T$, 
composed of a constant cost $c_0$ and variable costs
$C_v(\theta_{\tau},\tau)$ depending on the volume of damaged trees.
 
{The major differences with the model of \cite{Reed} are located in
  the expression of the net economic return $\mathcal Y$, more
  precisely in the thinning income and the final income in case of
  storm. In agreement with what \cite{Reed} rightly provided, if we
  do not take into account thinning incomes, the salvageable income in
  case of event is proportional to the total income in the absence of
  event. At the opposite in this paper, it is important to note that
  thinning incomes are included in total income. Thus, even if the
  final income in case of storm is proportional to the final income in
  the absence of risk, due the contribution of the thinning income,
  the total income in case of storm may not be proportional to the
  total income in the absence of storm.}

As the storm occurs independently of one another and randomly in time,
we deduce the following expression for the Faustmann Value:
\begin{align}
\label{W}
J_{\lambda} = & E(\sum_{i=1}^{\infty} e^{-\delta({\tau}_1+{\tau}_2+...+{\tau}_i)} \mathcal Y_i)
= \sum_{i=1}^{\infty} \prod_{j=1}^{i-1} E(e^{-\delta {\tau}_ j}).
E(e^{-\delta {\tau}_i} \mathcal Y_i)= {E(e^{-\delta \tau} \mathcal Y) \over 1 -
  E(e^{-\delta \tau})} 
\end{align}
Hence, the Faustmann Value is the expected value of the actualized sum
at the initial time of the net economic return $\mathcal Y$ and the
Faustmann Value:
\begin{align}
J_{\lambda} = E
[e^{-\delta \tau} (\mathcal Y+J_{\lambda})]
\end{align}
{ This recursive formulation shows that, contrary to \cite{Meilby},
  the land value is endogenous in our framework. Moreover this
  formulation can be used to facilitate the obtention of the Faustmann
  Value for more complex scenario.}

Let $E_{\mathcal V}(\tau)=E[\mathcal V(\theta_\tau,\xi_\tau,\tau)]$
the expected final income with respect to $\theta_{\tau}$ and
$\xi_{\tau}$ and $E_{C_v}(\tau) = E(C_v(\theta_\tau,\tau))$ the
expected variable clearing costs in case of a storm at time $\tau$,
$E_{R_k}=E[R(p_{u_k},u_k)]$ is the expected potential thinning income
at time $u_k$. Then, from Eq. (\ref{W}), we can deduce the Faustmann Value
(Appendix B.2-4):
\begin{align}
 J_{\lambda}= { E_W(0,T)- c_1-(c_0+c_1) a(T) \over b(T)} 
\end{align}
 where $\ds (\delta+\lambda) a(T) = \lambda
 (e^{(\delta+\lambda)(T-t_L)} -1)$, $ \ds b(T) =
 e^{(\delta+\lambda)T-\lambda t_L} - a(T)-1$ and $E_W(0,T)$ is the
 following modified expected income:
\begin{align}
\label{V}
  E_W(0,T) = & \sum_{k=1}^N \beta_{\delta,\lambda}^k {\bf \  ^tE_{R_k} .h_k} 
  +\lambda \int_{t_L}^T [E_{\mathcal V}(\tau)-E_{C_v}(\tau)]
  e^{(\delta+\lambda)(T-\tau)}d \tau + V(T) 
\end{align}

with $\beta^k_{\delta,\lambda} =
e^{(\delta+\lambda)(T-u_k)+\lambda(u_k-t_L)_-}$ the coefficient
relative to the thinning income at time $u_k$.

 As in the case without risk, the Faustmann Value can be deduced from
 an income $ E_W(0,T)$.  $E_W(0,T)$ is a modified linear combination
 of thinning incomes and final income.

  The major differences are the following: {
  first the discount rate $\delta$ is replaced by the addition of the
  discount rate and expected number of storms $\delta+\lambda$ (as
  previously noticed \cite{Reed} for the case without thinnings), 
  secondly the total income $W(0,T)$ is replaced by the modified
  expected income $E_W(0,T)$. More precisely, the thinning incomes
  $R_k$ are replaced by the expected thinning incomes $E_{R_k}$, the
  coefficients of the thinning income $\beta_{\delta,0}^k$ are
  replaced by $\beta_{\delta,\lambda}^k$ and finally an integral
  relative to the final cutting, minus clearing costs in case of a
  storm before the cutting age $T$ is added}.  We notice that the
coefficients of the thinning incomes $\beta^k$ only depend on risk and
economic parameters but are independent of the forest stand.

Moreover, if we assume that
the potential income $R(p,t)$ is linear in its first argument $p$, we
deduce (see Appendix B.5): $\ds E_{R_k} = R((1-E_{\rho_k})p_0,u_k)$
where $ E_{\rho_k}$ is the expectation of $\rho_{\sigma_k}(u_k) $ with
respect to the date of the last storm $\sigma_k$ before $u_k$.
In this case, the impact of risk of storm results in the substitution
of the reference price $p_0$ by the modified value $(1-E_{\rho_k})p_0$
for thinning time $u_k$.

{ The influence of forest growth and storm damage
  appears 
  in $E_{R_k}, E_{\mathcal V}(\tau),$ $E_{C_v}(\tau)$ and $V(T)$. This
  structure does not depend on the tree species.}

\subsection{The Expected Long-Run Average Yield}

The Expected Long-Run Average Yield is the expected net economic
return per unit of time ($\ds \lim_{t \rightarrow \infty}
E(\sum_{i=1}^{N_t} \mathcal Y_i)/t$ where $N_t$ is the number of
rotations before time $t$). Following the reasoning of \cite{Reed},
the Expected Long-Run Average Yield is equal to the ratio of the
averaged net economic return (stumpage minus regeneration and the
clearing costs in presence of risk) and the averaged effective cutting
age:
$\ds \overline Y_{\lambda} = E(\mathcal Y) / E(\tau) $, hence:
\begin{align}
     \overline Y_{\lambda} = \lambda {E_{W_0}(0,T) -c_1-(c_0+c_1)
  (e^{\lambda(T-t_L)}-1) \over \ds (1+\lambda t_L) e^{\lambda(T-t_L)} -1} 
\end{align}
 where: $\ds E_{W_0}(0,T)= \sum_{k=1}^N \beta_{0,\lambda}^k
{\bf \ ^tE_{R_k} .h_k} \ + \lambda \int_{t_L}^T [ E_{\mathcal
  V}(\tau)-E_{C_v}(\tau)] e^{\lambda(T-\tau)} d\tau +V(T)$

\subsection{Maximization of the criteria}

We consider the maximization of the Faustmann Value with respect to
thinnings and cutting age: $\ds \max_{(h_k)_k,(u_k)_k,T}J_{\lambda}$.
As in the case without risk, for fixed cutting age $T$, the
maximization of the Faustmann Value $\ds J_{\lambda}$ is equivalent to
the maximization of $E_W(0,T)$. Hence the maximization of the
Faustmann Value with respect to thinnings and cutting age
can be decomposed into two levels: $\ds \max_T [E_W^*(0,T)-
 c_1-(c_0+c_1) a(T)]/ b(T)$ where
$E_W^*(0,T)$ is the maximum of $E_W(0,T) $ with respect to thinnings:
$\ds \max_{(h_k)_k,(u_k)_k} E_W(0,T)$.  

We notice that for fixed thinnings $u_k, h_k$ and cutting age $T$,
knowing potential thinning income $R_k, E_{R_k}$, final income $V(T),
E_{\mathcal V}(t)$ and costs $E_{C_v}(t)$ for $t_L < t < T$ is
sufficient to express the Faustmann Value without or with risk.
Hence, from a numerical point of view, thanks to the analytical
obtained expressions of the criteria, a unique simulation of the stand
growth is needed to express the Faustmann Value and the Expected
Long-Run Average Yield without or with risk, for various discount
rates or function prices.

For fixed cutting age $T$, the behavior of the thinning income
$\beta^k$ coefficients will be studied to deduce the influence of
discount rate, risk and considered criteria (Faustmann Value, Expected
Long Run Average Value) in terms of optimal thinnings.

\subsubsection*{Dependence of optimal thinning with respect to discount
  rate}

The derivative of $\beta_{\delta,\lambda}^k$ with respect to the
discount rate $\delta$ normalized by $\beta_{0,\lambda}^k$ is:
\begin{align}
{1 \over \beta_{0,\lambda}^k} {\partial \beta_{\delta,\lambda}^k \over \partial \delta}|_{\delta
  =0}= T-u_k
\end{align}
We then deduce that the relative additional contribution of
$\beta_{\delta,\lambda}^k$ decreases as $k$ increases. Hence as in the
case without risk, for fixed cutting age $T$, the greater the discount
rate $\delta$, the earlier the optimal thinnings. { Moreover, for a
  sufficiently small risk rate $\lambda$, by continuity with respect
  to $\lambda$ we deduce that for fixed thinnings, the optimal cutting
  age decreases with respect to discount rate. Hence, also for the
  optimal cutting age, as in the case without risk, the greater the
  discount rate $\delta$, the earlier the optimal thinnings}.

\subsubsection*{Dependence of optimal thinning with respect to risk}

The derivative of $\beta_{\delta,\lambda}^k$ with respect to $\lambda$
normalized by $\beta_{\delta,0}^k$ is:
\begin{align}
  \label{premus} {1 \over \beta_{\delta,0}^k} {\partial
    \beta_{\delta,\lambda}^k \over \partial \lambda}|_{\lambda =0}= &
\begin{cases}
  T-t_L & \mbox{ if } u_k < t_L  \\
  T-u_k & \mbox{ if } u_k > t_L 
\end{cases}
\end{align}
We then deduce that the relative additional contribution of $\ds
\beta_{\delta,\lambda}^k$ decreases as $k$ increases. These results
imply that the thinnings at dates $u_k$ prior to $t_L$ have a greater
influence and the last thinnings a lower influence in case of risk. It
is not surprising because after $t_L$ the dates $u_k$ are less and
less reached as $k$ increases.  Hence, assuming that the integral term
in case of risk is neglectable, we deduce that, in presence of storm
risk, the optimal thinnings are done earlier than in the case without
risk for fixed cutting age.  { Moreover, for sufficiently small risk
  rate $\lambda$, it is possible, as for the $\delta$ dependence, to
  deduce that for fixed thinnings, the optimal cutting age decreases
  with respect to the risk rate $\lambda$.  So, for the optimal
  cutting age, the greater the risk rate $\lambda$, the earlier the
  optimal thinnings.}
The obtained result is generic and does not depend on the forest growth
nor on the species.
{ Our result is in accordance with forest managers decisions: shorter
  cutting age and earlier thinnings. Concerning the thinnings, our
  study enhances the previous argument concerning the $H/d$ ratio.}

\subsubsection*{An adjusted discount rate without risk compared to the
  risk case}

\cite{Reed} showed that in case of risk, without taking thinnings into
account, the optimal cutting age can be achieved by replacing $\delta$
by $\delta+\lambda$ in the case without risk. { What happens if we
  take thinnings into account ?} Without risk for an adjusted discount
rate $\delta+\lambda$, the derivative with respect to $\lambda$
normalized by $\beta_{\delta,0}^k$ is:
\begin{align}
\label{adjusted}
 {1 \over \beta_{\delta,0}^k} {\partial
    \beta_{\delta+\lambda,0}^k \over \partial \lambda}|_{\lambda =0}=
  &
  T-u_k
\end{align} { From the comparison of Eqs. (\ref{premus}) and
  (\ref{adjusted}) we first deduce that for $u_k < t_L$ the normalized
  derivatives of the coefficient are greater for an adjusted discount
  rate without risk than for the risk case: with the adjusted discount
  rate, the first thinnings are favoured and have a greater influence
  than in the risk case. Hence, assuming that the integral term in
  case of risk is neglectable, the optimal thinnings would be earlier
  for the adjusted discount rate without risk than for the real
  discount rate with risk. Secondly for $u_k > t_L$ the derivatives
  are equal. Hence, the optimal thinnings would be close to each other
  in this case. Moreover, for a sufficiently large value of the
  discount rate, the first optimal thinnings should satisfy $u_k <
  t_L$. At the opposite for sufficiently small value of discount rate
  the first optimal thinnings should satisfy $u_k > t_L$. We suggest
  the following conjectures:}

\begin{tabular}{ll}
  $(\mathcal C_{1})$:
  & If the first optimal thinnings for
  an adjusted discount rate  $\delta+\lambda$ without risk are \\
  & smaller than $t_L$ (potentially satisfied if $\delta+\lambda$ is not too small)
  then the optimal   \\
  & thinnings in an ajusted $\delta+\lambda$ without risk occur
  earlier than in the case  \\
  & with risk and discount rate $\delta$.
\end{tabular}

\begin{tabular}{ll}
  $(\mathcal C_{2})$: & 
  If the first optimal thinnings with an adjusted discount rate 
  $\delta+\lambda$ without risk are \\
  &  greater than $t_L$ (satisfied if $\delta+\lambda$ sufficiently small)
  then the optimal thinnings \\
  &  with an adjusted discount rate $\delta+\lambda$ without risk
  and with discount rate $\delta$ with  \\
  &  risk are close  to each other.
 \end{tabular}

 Taking $\delta=0$ in Conjecture $C_{2}$, we obtain that, for
 $\lambda$ not too high, the optimal thinnings for the Faustmann Value
 with discount rate $\lambda$ without risk and for the Expected
 Long-Run Average Yield with risk are close to each other.

 { Some information on the optimal thinnings dates with risk can be
   deduced from the comparison of the first optimal thinnings dates in
   the without risk case, using an adjusted discount rate, with
   $t_L$.}  The date $t_L$ is a threshold which permits to predict the
 behavior of the optimal thinnings with respect to the criteria. This
 conjecture has to be supported by numerical optimization in concrete
 cases.
\section{Application for a beech stand}

{ We now apply the analytic expressions given above to evaluate and
  compare the criteria without or with risk for a forest stand using a
  specific growth model. Maximization of the various criteria will
  permit in a first step to confirm or not the previous properties and
  conjectures.  In a second step we analyze the impact of the rate
  depreciation of timber price on silviculture}.
\subsection{Growth model}
We consider an average tree model, the state variables are the
tree-density $ n $ per hectar and 
the tree-basal area $s$ measured at breast $ 1.30$ m from the ground.
For initial tree-number $n(0)$ and tree-basal area $s(0)$ set, the
evolutions of $n$ and $s$ are governed by the following system of
equations:
\begin{eqnarray*}
 (\mathcal S_0) \ \ \ \ \ \ \ \ \ \ \ \ &
\begin{cases}
\ds  {dn(t) \over dt} &= -m(t) n(t) \\
\ds  {d s(t) \over dt} &= \ds {g(n(t),s(t)) \over n(t)} \Gamma(t) 
\end{cases}  \ \ \ \ \ \ \ \ \ \ \ \ \ \ \ \ \ \ \ \ \ \ \\
\mbox{ at the thinning times } u_k:&
\ n(u_k^+) = (1-h_k) n(u_k^-),\  0 \leq h_k < \overline h \notag
\end{eqnarray*}
where $m(.)$ is the natural mortality and { $g(n,s) \Gamma(t)$ is the
  instantaneous increase of the stand basal area:
  $\Gamma(t)$ is the potential increase at its peaks of tree-density
  and $g(n,s)$ a reductor depending on the tree-density.}

The thinning rates are limited to an upper value $\overline h$ {to
  insure regular income}.

In case of storm risk, the system $(\mathcal S_{1})$ to be considered
consists of equations of system $({\mathcal S_{0}})$, plus equation
(\ref{price}).
To complete the model,
the tree-height $H$ is expressed via a time-dependent function: $H(t)$
and the tree-volume $v(s,H)$ is a combination of the tree-basal area
and the tree height.

The expressions of the expected thinnings income and clearing costs if
a storm occurs are given in Appendix C.

\subsection{Optimization}

{ We are interested in the silviculture of a Beech stand: we give in
  Appendix D the related parameters and functions used for the
  evaluation of the criteria.}  As usual, the silviculture is
decomposed into two steps: a first step of respacings and a second
step of thinnings. We assume two respacings at time $10$ years and
$u_0=20$ years, the tree number is significantly reduced, the fixed
rate is $h_0=29/30$.  For the second step, the thinning dates are
separated by a minimum of $\Delta t$ years: $u_k-u_{k-1} \geq \Delta
t$ years. We choose $N=5, \Delta t = 8 $ years, $\overline h = .25$
year$^{-1}$ and we set the final tree-number $n_T=125$ stems/ha {in
  order to facilitate the comparison for various criteria}. To
summarize we determine, by optimization of the various criteria ($J_0,
\overline Y_0$ without and $ J_{\lambda}, \overline Y_{\lambda}$ with
risk), the optimal thinning dates $u_1,..,u_N$ and thinning rates
$h_1,..,h_N$ subject to the constraints:
\begin{align}
 & u_k-u_{k-1} \geq \Delta t ,& k = 1..N \notag \\
 & 0 \leq h_k \leq \overline h, & k = 1..N \notag \\
& n(T) = n_T & \notag
\end{align} 
The presented results are obtained by using the classical Nelder Mead
algorithm to optimize the various considered criteria.  As this
algorithm does not manage constraints on control variables, the
constraints on thinnings dates ($u_k-u_{k-1} \geq \Delta t$ years) and
final tree number ($n(T)=n_T$) are managed by using artificial
variables.

\section{Results and Discussion}

We analyze the optimal silviculture for the two criteria: Faustmann
Value (Table 1) and the Expected Long-Run Average Yield (Table 2).
For the two criteria, we give the optimal thinnings and cutting age
without and with risk: the thinnings dates $u_k$ in the first row and
the thinnings rate $h_k$ in the second row. For the Faustmann Value in
Table 1, we also give the scenario without risk, with a adjusted
discount rate $\delta+\lambda$ instead of $\delta$.
Finally in each table, the scenario with risk is given in the last
rows.  In all cases, we specify the expected effective values of the
final cutting age in parentheses (given by $\ds E(\tau) =
t_L+F(T-t_L)/ \lambda$).

Concerning the dependence of optimal thinning with respect to risk,
for the Faustmann value in Table 1, the first thinnings occur at
$71.8$ years without risk and $59.1$ years with risk.
For the Expected Long-Run Average Yield in Table 2, the difference in
the first thinnings dates is greater: $138.3$ years without risk and
$71.5$ years with risk. Hence for the two criteria, the optimal
thinnings occur earlier in the case with risk.
Moreover for the two criteria, the presence of risk implies earlier
thinnings and a shorter cutting age $T$.  We therefore make the risk
endogenous through optimization.  An earlier cutting age with risk
implies a lower final diameter at the cutting age.

Concerning the conjecture $(\mathcal C_{1})$, we compare the results
without risk obtained with $\delta+\lambda$ in the second scenario and
the result with risk obtained with $\delta$ in the third scenario of
Table 1. The results show that, in case of storm risk, by taking
thinnings into account, the optimal cutting ages are identical: $83.3$
years. It is in accordance with the previous result by \cite{Price}.
The first thinning date with the adjusted discount rate $50.9$ years
is smaller then $t_L=61.5$ years. In accordance with our conjecture,
the thinnings occur slightly earlier for the adjusted discount rate
without risk: $50.9$ years with the adjusted discount rate and $59.1$
years with the risk case for the respective first thinning dates.
Replacing $\delta$ by $\delta + \lambda$ in the case without risk to
mimic the case with risk is valid to predict the optimal age, { we
  do not obtain the optimal thinning dates but may predict the behavior
  of predicted thinning dates.}  Considering the conjecture $(\mathcal
C_{2})$ with $\delta=0$, first of all the cutting ages for $J_0$ and
$\overline Y_{\lambda}$ are respectively $105.3$ and $95.7$ years and
are of the same order of magnitude.
Secondly, the first thinning date, with discount rate $\lambda$,
$71.8$ years is later then $t_L=61.5$ years.  In accordance with this
conjecture,
the dates of the first thinnings are very similar: $71.8$ years for
$J_0$ and $71.5$ years for $\overline Y_{\lambda}$. A similar result
is found for the magnitudes.

In Tables 3 and 4, we present the optimized silviculture for two
values of rate depreciation of timber price $\alpha_d = .5,.9$. 
The real value of $\alpha_d$ is stochastic but in the range
$[.5,.9]$. From the results we deduce that, the greater the rate
depreciation $\alpha_d$, the earlier the optimal thinnings dates and
cutting age. Moreover there is no significant difference for the
tree-diameter.  { Finally, we remark that, in all optimizations, after
  the first thinning, the other thinnings are separated by the miminal
  authorized delay: $8$ years.}
\section{Conclusion}

We have studied the management of a stand in the presence of risk of
storm. From the analytic expressions of the criteria, we highlighted
the influence of the presence of storm risk on optimal thinning and
cutting age: in presence of storm risk, the optimal thinnings occur
earlier than in the case without risk. { Moreover, we highlight the
  role of a time threshold in the behavior of optimal thinnings with
  respect to the criteria. In particular we determine in which case it
  is valid or not to use an adjusted discount rate without risk to
  mimic the storm risk case. Simulations based on an average tree
  model for a beech stand validate the conjectures.}  More precisely,
if the first optimal thinnings with and adjusted discount rate and
without risk occur before the time threshold, the optimal thinnings
with an adjusted discount rate without risk are done earlier than in
the case with risk for the Faustmann Value.  At the opposite, if the
first optimal thinnings with and adjusted discount rate and without
risk are beyond the time threshold, the optimal thinnings and the
optimal rotation period for the Faustmann Value $J_0$ calculated with
discount rate $\lambda$ and for the Expected Long-Run Average Yield
are close to each other.

It can be observed that some forest managers want to drastically
reduce the cutting ages. Considering the second conjecture with
$\delta=0$, this can be interpreted in two ways. This desire can be
interpreted as an adaptation to climate change by taking into account
an increased risk of storm related to climate change. But it can also
be explained by the desire to change the criterion to maximize:
maximizing the Faustmann criterion may be preferred to maximizing the
Expected Long-Run Average Yield.

We showed that the obtention of analytic expressions for the criteria
opens the way to the analysis of the qualitative behavior of the
optimal thinnings and cutting age. It would be interesting to apply this
technique in a more complex risk framework especially for the
combination of consecutive various type of risks such as storm
followed by an epidemic event.

{\bf Acknowledgements}

The author would like to thank the anonymous reviewer whose comments
and suggestions helped us to improve the quality of this paper.

\begin{appendix}

\section*{Appendix A}

For fixed thinnings, omitting $T$ dependancy in $\mathcal H_0(0,T)$,
differentiating $J_0$ with respect to $T$ gives the following
first-order condition:
$$J_T= {(\delta \mathcal H_0 +V'(T))
  (e^{\delta T}-1)-\delta e^{\delta T} (H_0 + V(T)-c_1) \over
  (e^{\delta T}-1)^2} =0$$ Differentiating $J_T=0$ with respect to
$\delta$ yields: $\ds J_{TT} T_{\delta} + J_{T \delta}=0$. Hence, from
$\ds J_{TT} < 0$, we deduce that $T_{\delta}$ and $J_{T \delta}$ have
the same sign. 

If we let $\ds \mathcal H_1= {\partial \mathcal H_0 \over \partial
  \delta} = \sum_{k=1}^N {\bf \ ^tR_k.h_k} (T-u_k) e^{\delta
  (T-u_k)}$, $J_{T \delta}$ is proportional (with the same sign) to:
\begin{align}
  & (\mathcal H_0+\delta \mathcal H_1)(e^{\delta T}-1) +(\delta
  \mathcal H_0+V'(T))T e^{\delta T}-(1+\delta T)
  e^{\delta T}(\mathcal H_0+V(T)-c_1)-\delta e^{\delta T} \mathcal H_1 \notag \\
  & (\mbox{ then using } J_T =0) \notag \\
  = & (\mathcal H_0+\delta \mathcal H_1)(e^{\delta T}-1)-\delta
  e^{\delta T} \mathcal H_1 +[{\delta T e^{\delta T} \over e^{\delta
      T}-1} -(1+\delta T)] (\mathcal H_0+V(T)-c_1) e^{\delta T} \notag \\
  = & \mathcal H_0 (e^{\delta T}-1) -\delta \mathcal H_1 +(1+\delta
  T-e^{\delta T})
  {e^{\delta T} \over e^{\delta T}-1}(\mathcal H_0+V(T)-c_1) \notag \\
  & \mbox{ which is proportional, with the same sign, to }: \notag \\
  = & ((\delta T-1) e^{\delta T}+1)\mathcal H_0 -\delta(e^{\delta
    T}-1) \mathcal H_1
  -(e^{\delta T}-1-\delta T)e^{\delta T} (V(T)-c_1) \notag \\
  & (\mbox{ then, from the expressions of } \mathcal H_0, \ \mathcal H_1): \notag \\
  = & \sum_{k=1}^N \ R_k.h_k [((\delta T-1) e^{\delta T}+1)
  -\delta(T-u_k)(e^{\delta T}-1)]e^{\delta(T-u_k)} -(e^{\delta
    T}-1-\delta T)e^{\delta T} (V(T)-c_1)  \notag \\
  = & \sum_{k=1}^N \ R_k.h_k [\delta u_k(e^{\delta T}-1) -(e^{\delta
    T}-1-\delta T)]e^{\delta(T-u_k)} -(e^{\delta T}-1-\delta
  T)e^{\delta T} (V(T)-c_1) \notag
\end{align}

We remark that $R_k$ is increasing with respect to $k$, hence the
major contribution to the sum is the last thinning. Moreover for small
value of $\delta$, the coefficients of $R_k h_k$ for $u_k \sim T$ is
closed to $\ds e^{\delta T}-1-\delta T$, hence for sufficiently small
value of $h_k$ and $c_1$, we deduce that $J_{T \delta}$ is negative.

\section*{Appendix B}

\subsection*{B.1 Evaluation of $\mathcal H(t_1,t_2,\sigma)$}

Let $\sigma=(\sigma_k)_{k \in K}$ the vector composed of
$\sigma_k=\tau^-_{u_k}$ the date of the last storm before $u_k$ for
each $k \in K=\{ k| u_k < t_L+\Delta \}$. We denote
$\mathcal H(t_1,t_2,\sigma)$ the thinning income between $t_1$ and
$t_2$, taking into account the dates of the last storm before thinning
times. Hence the modified value of the timber price reference is
$p_{u_k}=(1-\rho_{\sigma_k}(u_k))p_0$, and $\mathcal
H(t_1,t_2,\sigma)$ can be deduced:
$$ \mathcal H(t_1,t_2,\sigma) = \sum_{t_1 < u_k \leq t_2}
{\bf \ ^tR((1-\rho_{\sigma_k}(u_k))p_0,u_k) .h_k} \  e^{\delta (t_2-u_k)} $$

\subsection*{B.2 Evaluation of $E(e^{-\delta \tau})$}

We express $\ds E(e^{-\delta \tau})=
\int_{t_L}^{T} e^{-\delta \tau} dF(\tau-t_L) + e^{-\delta T}
(1-F(T-t_L))= {\lambda + \delta e^{-(\delta+\lambda)(T-t_L)} \over
  \delta +\lambda} e^{-\delta t_L}$

\subsection*{B.3 Evaluation of $E(-e^{\delta \tau} \mathcal Y)$}

Let $E_{R_k}=E[R((1-\rho_{\sigma_k}(u_k))p_0,u_k)]$ the expected
potential thinning income at time $u_k$, then the expected thinning
income between $t_1$ and $t_2$ is::
$$ E_{\mathcal H}(t_1,t_2) = E_{\sigma}[\mathcal H(t_1,t_2,\sigma)] =
\sum_{t_1 < u_k \leq t_2} {\bf \ ^tE_{R_k} .h_k} \ e^{\delta
  (t_2-u_k)}$$
Knowing the expressions of the expected incomes and costs:
\begin{align}
  E(e^{-\delta \tau} \mathcal Y)=& E_{\mathcal H}(0,t_L) e^{-\delta
    t_L} +\int_{t_L}^T (E_{\mathcal H}(t_L,\tau) +{E_{\mathcal V}}(\tau) -c_1
  -c_0-E_{C_v}(\tau)) e^{-\delta \tau}
  dF(\tau-t_L) \notag \\
  & + (E_{\mathcal H}(t_L,T)+V(T)-c_1)e^{-\delta T} (1-F(T-t_L))
  \notag
\end{align}
The part relative to the thinning after $t_L$ in $E(e^{-\delta \tau}
\mathcal Y)$, by inverting the integral and the summation, is:
\begin{align}
  \int_{t_L}^T E_{\mathcal H}(t_L,\tau)
e^{-\delta    \tau} dF(\tau-t_L) + E_{\mathcal H}(t_L,T)
e^{-\delta T}  (1-F(T-t_L))  = 
  \sum_{t_L < u_k < T} {\bf \ ^tE_{R_k} .h_k} \ e^{-\delta u_k-\lambda(u_k-t_L)}
 \notag
\end{align}
The part in $E(e^{-\delta \tau} \mathcal Y)$ relative to 
 the clearing costs is:
$$\int_{t_L}^T (c_0+E_{C_v}(\tau)) e^{-\delta \tau} dF(\tau-t_L) 
= c_0{\lambda \over \delta+\lambda}
(1-e^{-(\delta+\lambda)(T-t_L)}) e^{-\delta t_L}+ \int_{t_L}^T
E_{C_v}(\tau) e^{-\delta \tau} dF(\tau-t_L)$$

Finally:
\begin{align}
  E(e^{-\delta \tau} \mathcal Y) = & \sum_{k=1}^N {\bf \
    ^tE_{R_k} .h_k} \ e^{-\delta u_k-\lambda (u_k-t_L)_+} + \lambda
  \int_{t_L}^T [E_{\mathcal V}(\tau)-E_{C_v}(\tau)]
  e^{-\delta \tau-\lambda(\tau-t_L)} d\tau \notag \\
  & +(V(T)-c_1)e^{-\delta T-\lambda (T-t_L)} - (c_0+c_1){\lambda \over
    \delta+\lambda} (1-e^{-(\delta+\lambda) (T-t_L)})e^{-\delta t_L}
  \notag
\end{align}

\subsection*{B.4 The Faustmann Value}

By using the expression of the expectations $E(e^{\delta \tau})$,
$E(e^{\delta \tau} \mathcal Y)$, multiplying the numerator and the
denominator of (\ref{W}) by $\ds e^{(\delta+\lambda)(T-t_L)+\delta
  t_L}$ and reordering the terms, we obtain the Faustmann Value
$J_{\lambda}$:
\begin{align}
J_{\lambda}= & {E_W(0,T)-c_1 -
  (c_0+c_1){\lambda \over \delta+\lambda}
  (e^{(\delta+\lambda)(T-t_L)}-1) \over
  e^{(\delta+\lambda)(T-t_L)+\delta t_L} - {\lambda \over \delta
    +\lambda} e^{(\delta+\lambda)(T-t_L)} - {\delta \over \delta
    +\lambda} } \notag
\end{align}
where $E_W(0,T)$ is the modified expected income (\ref{V}). 

\subsection*{B.5 Evaluation of $E_{R_k}$}

From the assumption that $R$ is linear in its first argument: $\ds
E_{R_k} = E(R((1-\rho_{\sigma_k}(u_k))p_0,u_k))=
R((1-E_{\rho_k})p_0,u_k)$ where $\ds E_{\rho_k}=
E_{\sigma_k}[\rho_{\sigma_k}(u_k)] $.
For a sufficiently small $\Delta$, we obtain:
\begin{align}
  E_{\rho_k}= & I_{u_k > \Delta} E_{\sigma_k}[\rho_{\sigma_k}(u_k)
  I_{u_k-\Delta < \sigma_k < \min(u_k,t_L)}]
  = I_{u_k > \Delta} E_{\sigma_k}[\rho_0(u_k-\sigma_k) I_{\max(u_k-t_L,0)
    < u_k-\sigma_k < \Delta}] \notag \\
  =& I_{\Delta < u_k < t_L+\Delta} \int_{\max(u_k-t_L,0)}^{\Delta}
  \rho_0(x) dF(x) \notag
\end{align}

\section*{Appendix C: Expected incomes and costs: $E_{\mathcal V}$ and
  $E_{C_v}$}

Assuming that the timber price per unit of volume is the product of
the timber price reference $p_0$ and a quality related function $q$
depending on the tree basal-area $s$, the income function $R$:
$R(p_0,t) = [p_0 q(s(t))-e_0] v_T(t) $ where $v_T(t)$ is the total
volume $v_T(t)=v(s(t),H(t))n(t)$ and $e_0$ is the operating cost per
unit of volume. The quality return function $q$ is assumed increasing
with respect to the tree-basal area $s$.
 Due to the assumptions
made on $\theta$ and $\xi$, the final income $\mathcal V$ is
given by $ \ds \mathcal V(\theta,\xi,\tau)= [(1-\theta)(1-\xi) p_0
q(s(\tau))-(1-\theta)e_0] v_T(\tau)$.
We assume that the variable clearing costs $C_v$ linearly depend on
the volume of damaged trees $ \theta_{\tau} v_T(\tau)$ after a storm,
hence
$\ds C_v(\theta_{\tau},\tau)=c_v \theta_{\tau} v_T(\tau)$. We notice
that, if $e_0 > 0$, the final income in case of storm is proportional
to the final income without risk.  From the definition of $\alpha$ and
$\alpha_p$, we deduce the expected final income $E_{\mathcal V}(\tau) =
[\alpha_p(\tau) p_0 q(s(\tau)) - \alpha(\tau) e_0] v_T(\tau)$ and the
expected variable clearing costs $\ds E_{C_v}(\tau) =
c_v(1-\alpha(\tau)) v_T(\tau)$.  

\section*{Appendix D: Functions and parameters of the models}

\subsection*{D.1 The growth model}

The reductor is $\ds g(n,s) = 1-e^{-m_1 n \sqrt{s}}$ and the potential
increase $ \Gamma(t) = m_2+m_3 d H(t)/dt $ where the tree-height
$H(t)$ at time $t$ given by $\ds H(t)= H_0 \ds (1- e^{-m_4 t})$
(\cite{Dhote}) and $H_0$ the limited tree-height which depends mainly
on soil fertility.
The time unit for the case study is the year.  We assume usual values
for the initial tree number
$n(0)=8000$ stems/ha and the initial tree basal area $s(0) = .000125$
m$^2$. 

\subsection*{D.2 Risk modelling}

The expected number of storms per year is: $\lambda=0.01$ year$^{-1}
$.  We assume no damage under a height $H_L= 21$ m, a height limit
$H_0=40$ m, so from the expression of $H(t)$ we deduce $ t_L = -
\log(1-H_L / H_0)/m_4= 61.5$ years. From data provided in
\cite{Hanew}, we infer that the expected proportion of surviving trees
is: $\ds \alpha(t) = 1- \phi((H(t)-H_L)/(H_0 -H_L))$, with $\phi$ an
increasing function in $x$, (we choose $\phi(x)=\min(.1+1.5 x,1)
$). { Due to the lack of a precise model for beech stand, we do not
  take into account the $H/d$ dependence.}

\subsection*{D.3 Economic parameters and functions}

The timber price reference is $p_0= 57$ Euros/m$^3$. The timber price
quality related function depending on the tree basal area is $\ds
q(s)= 12/57+45/57 (1-(1-\sqrt{s / 0.3318})^{1.8})$ for $ 0.0038 < s <
0.3318$ m$^2$ which has been fitted from data in \cite{Tarp}. 
The operating cost per unit of volume is $e_0 = 10 $ Euros/m$^3$ and 
the regeneration cost is $c_1=1000$ Euros/ha.

Criteria are evaluated with the discount rate $\delta = 0.01$
year$^{-1}$. The rate depreciation of timber price following a storm
is $\alpha_d=.5$, the perturbation time is $\Delta=5$ years and
$ \rho_0(t) = \alpha_d (1-t^4 / \Delta^4)$ for $0 < t < \Delta$.

\end{appendix}

\pagebreak

\noindent {\bf Table 1}: Optimal Faustmann Value with respect to thinning rates
$h_k$, dates $u_k $ and cutting age $T$.

\noindent \begin{tabular}{l|cccccccc|c} \hline
  Scenario &\multicolumn{6}{c}{Optimal Thinnings} & Cutting age  & Diameter & Faustmann Value   \\
 \hspace{8mm} discount rate  & \multicolumn{6}{c}{ (year)}& (year)& (cm) & (Euro/ha)    \\
  \hline
  Without risk&&&& &&&& \\
  \hspace{8mm} $\delta$   &&&$71.8$ & $79.8$ &$87.9$&&{ $105.3$} & $56.4$&$16515$ \\
  & && $.186$&$.247$ &$.234$          & &&& \\
  &&&&& &&&& \\
  \hspace{8mm}$ \delta+\lambda$     &$50.9$ & $58.9$ &$67.0$&&&&$84.0$ & $53.2$ & $ 4725$ \\
   & $.246$&$.222$ &$.200$&&  &&&& \\

  &&& &&&&&& \\
  \hline
  With risk &&&&&&&& \\
  \hspace{8mm} $\delta$&& $59.1$ &$67.2$&$75.3$&&&$83.3(81.2)$&$50.6$&$13994$ \\
  &  & $.240$   &$.250$&$.177$&&& && \\
  \hline
\end{tabular}

\vspace{5mm}

\noindent {\bf Table 2}: Optimal LRAY with respect to thinning rates $h_k$,
dates $u_k $ and cutting age $T$.

\noindent \begin{tabular}{l|cccccccc|c}
  \hline
  Scenario &  \multicolumn{6}{c}{Thinnings} & Cutting age  & Diameter & LRAY \\ 
  &   \multicolumn{6}{c}{ (year)}& (year)&(cm) & (Euro/ha)    \\ 
  \hline
   Without risk&&&&&&&&& \\
  \hspace{8mm}   & &&&$138.3$ &$146.4$ &$154.5$  &$162.8$ & $61.7$ &$ 312.8$ \\
  &  & &&$.167$ &$.250$ & $.250$               & && \\
  \hline
         With risk &&&&&&&& \\
  \hspace{8mm}  & $71.5$ &$79.5$&$87.6$ &&&&$95.7(90.5)$ & $ 52.6$ & $219.8$ \\
  &$.167$&$.250$&$.250$&& &&&&\\
  \hline
\end{tabular}

\vspace{5mm}
\noindent {\bf Table 3}: Optimal Faustmann Value for two values of rate depreciation.

\noindent \begin{tabular}{l|cccccccc|c}
  \hline
  Scenario &  \multicolumn{6}{c}{Thinnings} & Cutting age  & Diameter & Faustmann Value \\ 
  &   \multicolumn{6}{c}{ (year)}& (year)&(cm) & (Euro/ha)    \\ 
\hline
  $\alpha_d = .5$&& $59.1$ &$67.2$&$75.3$&&&$83.3(81.2)$&$50.6$&$13994$ \\
  &  & $.240$   &$.250$&$.177$&&& && \\
  $\alpha_d = .9$&& $56.3$ &$64.3$&$72.3$&&&$80.3(78.3)$&$49.5$&$13318$ \\
  &  & $.186$   &$.250$&$.231$&&& && \\
  \hline
\end{tabular}

\vspace{5mm}

\pagebreak

\noindent {\bf Table 4}: Optimal LRAY for two values of rate depreciation.

\noindent \begin{tabular}{l|cccccccc|c}
  \hline
  Scenario &  \multicolumn{6}{c}{Thinnings} & Cutting age  & Diameter & LRAY \\ 
  &   \multicolumn{6}{c}{ (year)}& (year)&(cm) & (Euro/ha)    \\ 
  \hline
   $\alpha_d = .5$ && $71.5$ &$79.5$&$87.6$ &&&$95.7(90.5)$ & $ 52.6$ & $219.8$ \\
  &&$.167$&$.250$&$.250$&& &&&\\
   $\alpha_d = .9$ && $65.6$ &$73.6$&$81.7$ &&&$89.7(86.1)$ & $ 51.9$ & $206.5$ \\
  &&$.249$&$.235$&$.182$&& &&&\\
   \hline
\end{tabular}

\end{document}